%% file: main.tex
\documentclass[letterpaper, 10 pt, conference]{ieeeconf}

\IEEEoverridecommandlockouts

\usepackage{balance}

\usepackage{amsmath,mathtools}  %
\usepackage{amsfonts,amssymb}
\usepackage{bbm}

\usepackage{xspace}    %

\usepackage{graphicx}
\graphicspath{{figs/}}

\usepackage{subcaption}

\usepackage{hyperref}
\hypersetup{colorlinks=true, linkcolor= blue, allcolors=blue, citecolor = blue, filecolor=black, urlcolor=blue}

\usepackage{pgfplots}\pgfplotsset{compat=1.9}
\usepackage{grffile}
\pgfplotsset{compat=newest}  %
\usetikzlibrary{plotmarks}
\usetikzlibrary{arrows.meta}
\usepgfplotslibrary{patchplots}

\pgfplotsset{compat=1.15}
\usepackage{mathrsfs}
\usetikzlibrary{arrows}

\usepackage{tabularx}

\usepackage[ruled]{algorithm}
\usepackage[noend]{algpseudocode}

\makeatletter
\newcommand{\multiline}[1]{%
  \begin{tabularx}{\dimexpr\linewidth-\ALG@thistlm}[t]{@{}X@{}}
    #1
  \end{tabularx}
}
\makeatother

\usepackage[per-mode=symbol]{siunitx}

\usepackage{cite}  %
\usepackage{array}  %
\usepackage{booktabs}           %

\newtheorem{assumption}{Assumption}

\newcounter{remark}
{\par\endtrivlist\unskip}

\newcounter{problem}
{\par\endtrivlist\unskip}

\usepackage[extramath,probability,reviewmark]{truonglatexdefs}

\input{defs}

\usepackage{stfloats}%

\title{\LARGE \bf
Optimal Weight Adaptation of Model Predictive Control for Connected and Automated Vehicles in Mixed Traffic with Bayesian Optimization}
\author{Viet-Anh Le, \emph{IEEE Student Member}, Andreas A. Malikopoulos, \emph{IEEE Senior Member}%
\thanks{This work was supported by NSF under Grants CNS-2149520 and CMMI-2219761.}
\thanks{
The authors are with the Department of Mechanical Engineering, University of Delaware, Newark, DE 19716 USA. E-mail: \tt\small{vietale@udel.edu}, \tt\small{andreas@udel.edu}.}
}

\begin{document}

\maketitle
\thispagestyle{empty}
\pagestyle{empty}

\begin{abstract}
In this paper, we develop an optimal weight adaptation strategy of model predictive control (MPC) for connected and automated vehicles (CAVs) in mixed traffic.
We model the interaction between a CAV and a human-driven vehicle (HDV) as a simultaneous game and formulate a game-theoretic MPC problem to find a Nash equilibrium of the game.
In the MPC problem, the weights in the HDV's objective function can be learned online using moving horizon inverse reinforcement learning.
Using Bayesian optimization, we propose a strategy to optimally adapt the weights in the CAV's objective function 
so that the expected true cost when using MPC in simulations can be minimized.
We validate the effectiveness of the optimal strategy by numerical simulations of a vehicle crossing example at an unsignalized intersection.
\end{abstract}

\section{Introduction}
\label{sec:intro}

Recent advancements in connected and automated vehicles (CAVs) provide a promising chance in reducing both energy consumption and travel delay \cite{guanetti2018control,ersal2020connected}.
In our previous work \cite{Malikopoulos2020,chalaki2020TCST,Bang2022combined}, we addressed coordination and routing problems for CAVs given full penetration of CAVs.
However, CAVs will gradually penetrate the market and co-exist with human-driven vehicles (HDVs) in the next decades.
Therefore, addressing safe and efficient motion planning and control for CAVs in mixed traffic given various human driving styles is highly important.
Several control approaches have been proposed in the literature such as %
model predictive control \cite{mahbub2022_ifac,wang2022data}, learning-based control \cite{chalaki2020ICCA,valiente2022robustness}, game-theoretic control \cite{chandra2022gameplan}, and socially-compatible control \cite{schwarting2019social,wang2021socially}.

Among those control approaches, model predictive control (MPC) has received significant attention since (1) it can be integrated into other methods such as learning-based control or socially-compatible control, and (2) it can handle multiple objectives and constraints concurrently.
However, like in many MPC approaches for dynamical systems, %
some objectives, constraints, or system dynamics in motion planning and control for CAVs are usually simplified or approximated so that the resulting MPC problems can be solved in real-time.
In addition, the objective function in MPC is generally formed by a linear combination of multiple features, in which the weights are chosen empirically.
As a result, true cost optimization might not be achieved leading to performance degradation if the weights are chosen inappropriately.
An efficient technique to overcome these difficulties in practice is automatic weight tuning \cite{hewing2020learning} which aims to derive a strategy to tune the weights of MPC so that the best true cost can be achieved. 
Marco \etal \cite{marco2016automatic} used Bayesian optimization to optimize weights of a cost function to compensate for the discrepancy between the true dynamics and a linearized model.
Gros and Zanon \cite{gros2019data} utilized reinforcement learning for parameter adaptation in nonlinear MPC.
Jain \etal \cite{jain2021optimal} focused on finding an MPC rollout having a low true cost using covariance matrix adaptation evolution strategy.

Furthermore, in the control applications involving human decisions, \eg CAVs interacting with HDVs in mixed traffic, the controller must address the stochasticity and diversity caused by human behavior.
Generally, MPC with fixed weights cannot guarantee to work well in such applications. %
For example, overly weighting toward the safety objective in the MPC design while encountering a driving scenario with a conservative HDV may cause traffic delay. %
In contrast, if CAVs and HDVs behave aggressively then unsafe situations may occur.
Therefore, the weights of the MPC problem need to be adapted online depending on the human driving model.

In the recent research effort \cite{Le2022cdc}, we developed a control framework to address the motion planning problem for CAVs in mixed traffic.
We modeled the interaction between a CAV and an HDV as a simultaneous game and 
proposed an MPC objective function to find a Nash equilibrium of the game.
The weights in the objective function are parameterized by social value orientation (SVO), and depending on the online estimate of the SVO for the HDV, the MPC weights are adapted heuristically.
In this paper, we propose a method for \emph{optimal weight adaptation} of MPC for CAVs in mixed traffic based on \emph{Bayesian optimization}.
Using the proposed method, we can derive offline the optimal weight adaptation strategy for the MPC with respect to the HDV's objective weights so that the true desired performance can be achieved. %
Then by learning the objective weights that best describe human driving behavior online using real-time data and the moving horizon inverse reinforcement learning (IRL) technique \cite{ziebart2008maximum}, 
the MPC weights are adapted accordingly.
We demonstrate the proposed method by a vehicle crossing example at an unsignalized intersection, and show the benefits by comparing with the heuristic method in \cite{Le2022cdc}.

The remainder of this paper is structured as follows.
Section~\ref{sec:mpc} presents the game-theoretic MPC formulation and the moving horizon IRL technique.
In Section~\ref{sec:algo}, we develop the method to derive the optimal weight adaptation strategy with Bayesian optimization.
In Section~\ref{sec:example}, we demonstrate the proposed framework by an intersection crossing example, while numerical simulation results are provided in Section~\ref{sec:sim}.
Finally, we conclude the paper in Section~\ref{sec:conc}.

\section{Motion Planning for CAVs in Mixed Traffic with Model Predictive Control}
\label{sec:mpc}
In this section, we present a game-theoretic MPC formulation for motion planning of a CAV while interacting with an HDV along with the moving horizon IRL technique to learn the objective weights of the HDV from real-time data. 

\subsection{Model Predictive Control for Motion Planning}

We consider an interactive driving scenario including a CAV and an HDV whose indices are $1$ and $2$, respectively.
The goal of the MPC motion planner is to generate the trajectory and control actions of \CAV{1} while considering the real-time driving behavior of \HDV{2}.
To guarantee that \CAV{1} has data of \HDV{2}'s real-time trajectories, we make the following assumption: 
\begin{assumption}
\label{assp:coordinator}
A coordinator is available to collect trajectories of \HDV{2} and transmit them to \CAV{1} without any significant delay or error during the communication.
\end{assumption}

We formulate the problem in the discrete-time domain, in which the dynamic model of each vehicle $i$ is given by
\begin{equation}
\label{eq:dynamic}
\bb{x}_{i,k+1} = \bb{f}_i (\bb{x}_{i,k}, \bb{u}_{i,k}) ,
\end{equation}
where $\bb{x}_{i,k}$ and $\bb{u}_{i,k}$, $i = 1,2$, are the vectors of states and control actions, respectively, at time step $k \in \NN$.
We utilize the control framework presented in \cite{Le2022cdc}, in which the interaction between \CAV{1} and \HDV{2} is modeled as a simultaneous game, \ie the game without a leader-follower structure, in which the objective of each vehicle includes its individual objective and a shared objective.
Let $l_{1} \big(\bb{x}_{1,k+1}, \bb{u}_{1,k})$ and $l_{2} \big(\bb{x}_{2,k+1}, \bb{u}_{2,k})$ be the individual objective functions of \CAV{1} and \HDV{2}, %
respectively, and $l_{12} \big(\bb{x}_{12,k+1}, \bb{u}_{12,k}\big)$, where $\bb{x}_{12,k+1} = [\bb{x}^\top_{1,k+1}, \bb{x}^\top_{2,k+1}]^\top$ and $\bb{u}_{12,k} = [\bb{u}^\top_{1,k}, \bb{u}^\top_{2,k}]^\top$, be the cooperative term at time step $k$.
We assume that \CAV{1} and \HDV{2} share the same cooperative objective, \eg collision avoidance.
Those objective functions are usually designed as weighted sums of some features as follows
\begin{align}
\label{eq:li}
l_{i} \big(\bb{x}_{i,k+1}, \bb{u}_{i,k}) &= \bbsym{\omega}_i^\top \bbsym{\phi}_i \big(\bb{x}_{i,k+1}, \bb{u}_{i,k}), \; i = 1, 2 , \\
\label{eq:l12}
l_{12} \big(\bb{x}_{12,k+1}, \bb{u}_{12,k}) &= \bbsym{\omega}_{12}^\top \bbsym{\phi}_{12} \big(\bb{x}_{12,k+1}, \bb{u}_{12,k}) ,
\end{align}
where $\bbsym{\phi}_{i}$, $\bbsym{\phi}_{12}$ are vectors of features and $\bbsym{\omega}_{i} \in \WWW_i$, $\bbsym{\omega}_{12} \in \WWW_{12}$ are corresponding vectors of weights, where $\WWW_i$ and $\WWW_{12}$ are the sets of feasible values.
For ease of notation, we define $-i$ for each $i \in \{1,2\}$ as the other vehicle than vehicle $i$. 
We consider that given any control actions $\bb{u}_{-i,k}$ of the other vehicle, each vehicle $i$ applies the control actions $\bb{u}_{i,k}^*$ that minimizes a sum of its individual objective and the shared objective, \ie
\begin{equation}
\label{eq:utility}
\bb{u}_{i,k}^* \hrm{2} = \hrm{2}
\underset{\bb{u}_{i,k}} \argmin \, l_i \big(\bb{x}_{i,k+1}, \bb{u}_{i,k}) + l_{12} \big(\bb{x}_{12,k+1}, \bb{u}_{12,k}), \, \forall \bb{u}_{-i,k}.
\end{equation}

Next, we formulate an MPC problem with a control horizon of length $H \in \NN$.
Let $t$ be the current time step
and $\III_t = \{ t, \dots, t+H-1 \}$ be the set of all time steps in the control horizon at time step $t$.
We can recast the simultaneous game between \CAV{1} and \HDV{2} presented above as a potential game \cite{marden2009cooperative}, the game in which all players minimize a single global function called the potential function.
In the potential game, a Nash equilibrium can be found by minimizing the potential function.
The potential function in this game at each time step $k$ is
\begin{equation}
\begin{split}
& l_{\text{pot}} \big(\bb{x}_{12,k+1}, \bb{u}_{12,k}) \\
&\quad = \hrm{3} \sum_{i = 1,2} l_{i} \big(\bb{x}_{i,k+1}, \bb{u}_{i,k}) + l_{12} \big(\bb{x}_{12,k+1}, \bb{u}_{12,k}) %
\end{split}
\end{equation}
Therefore, we propose utilizing the cumulative sum of the potential function over the control horizon as the objective function in the MPC problem, which can be given by
\begin{equation}
\label{eq:mpc-obj-func}
J_{\text{MPC}} = \sum_{k \in \III_t} l_{\text{pot},k} \big(\bb{x}_{12,k+1}, \bb{u}_{12,k}).
\end{equation}

Hence, the MPC problem for motion planning of \CAV{1} is formulated as follows
\begin{subequations}
  \label{eq:mpc}
  \begin{align}
    &
    \begin{multlined}
    \underset{ \{\bb{u}_{12,k}\}_{k \in \III_t} }{\text{minimize}} \quad J_{\text{MPC}} %
    \end{multlined}
    \label{eq:mpc:obj}\\
    & \text{subject to:} \nonumber  \\
    & \qquad \text{\eqref{eq:dynamic}},\; \, i = 1,2, \label{eq:mpc-dyn} \\
    & \qquad g_j (\bb{x}_{12,k+1}, \bb{u}_{12,k}) \le 0,\, \forall j \in \JJJ_{\text{ieq}}, \label{eq:mpc-ineq} \\
    & \qquad h_j (\bb{x}_{12,k+1}, \bb{u}_{12,k}) = 0, \, \forall j \in \JJJ_{\text{eq}}, \label{eq:mpc-eq}
  \end{align}
\end{subequations}
where \eqref{eq:mpc-dyn}\textendash\eqref{eq:mpc-eq} hold for all $k \in \III_t$.
The constraints \eqref{eq:mpc-ineq} and \eqref{eq:mpc-eq} are inequality and equality constraints with $\JJJ_{\text{ieq}}$ and $\JJJ_{\text{eq}}$ are sets of indices.

In the objective function of the MPC problem \eqref{eq:mpc}, assume that we can pre-define the features $\bbsym{\phi}_i,\; i = 1,2$ and $\bbsym{\phi}_{12}$, if we learn online $\bbsym{\omega}_{2}$ and $\bbsym{\omega}_{12}$ that best describe the human driving behavior, 
the CAV's objective weights $\bbsym{\omega}_{1}$ are adapted to achieve the desired performance.
The optimal strategy for adapting $\bbsym{\omega}_{1}$ can be derived offline using Bayesian optimization as presented in Section~\ref{sec:algo}.

\subsection{Moving Horizon Inverse Reinforcement Learning}

To identify the weights $\bbsym{\omega}_2$ and $\bbsym{\omega}_{12}$ in the individual objective function of \HDV{2} and the shared objective, we utilize the feature-based IRL approach \cite{ziebart2008maximum,kuderer2015learning}, a machine learning technique developed to learn the underlying objective or reward of an agent by observing its behavior.
We define the vector of all features and the vector of all corresponding weights in \HDV{2}'s objective function as $\bb{f} = [\bbsym{\phi}^\top_{2}, \bbsym{\phi}^\top_{12}]^{\top}$ and $\bbsym{\theta} = [\bbsym{\omega}^\top_{2}, \bbsym{\omega}^\top_{12}]^{\top}$, respectively.
Let $\tilde{\bb{f}}$ be the vector of average observed feature values computed from data and $\EE_{p} [\bb{f}]$ be the expected feature values with a given probability distribution $p$ over trajectories.
With feature-based IRL, the goal is to learn the weight vector $\bbsym{\theta} \in \Omega$, where $\Omega = \WWW_2 \times \WWW_{12}$ so that expected feature values can match observed feature values.

In moving horizon IRL, at each time step, we utilize the $L \in \NN$ most recent trajectory segments to update the weight estimate, where $L$ is the estimation horizon length.
Let $t$ be the current time step and $\RRR_t = \{ \bb{r}_m \}_{m = 1,\dots, L}$ be the set of $L$ sample trajectory segments collected over the estimation horizon at time $t$, in which 
$\bb{r}_m = ( \bb{x}_{12,t-m}, \bb{x}_{12,t-m+1}, \bb{u}_{12,t-m} )$, for $m = 1,\dots, L$,  
is the tuple representing the trajectory segment.
We use the maximum entropy IRL approach \cite{ziebart2008maximum} that utilizes an exponential family distribution for $p$ and maximizes the entropy of the distribution, yielding the following optimization problem
\begin{equation}
\label{eq:opt-phi}
\underset{\bbsym{\theta} \in \Omega}{\maximize} \sum_{\bb{r}_m \in \RRR} \log \, p \big(\bb{r}_m \,|\, \bbsym{\theta} \big).
\end{equation}

To solve \eqref{eq:opt-phi}, one can use gradient-based methods where the gradient can be approximated by the difference between the expected and the empirical feature values \cite{ziebart2008maximum}
\begin{equation}
\label{eq:grad-f}
\nabla \LLL_{\bbsym{\theta}}
= \tilde{\bb{f}} - \EE_{p} [\bb{f}].
\end{equation}
The average observed feature values $\tilde{\bb{f}}$ can be computed from an average of feature values for all training samples
\begin{equation}
\label{eq:aver-fea}
\tilde{\bb{f}} = \frac{1}{L} \sum_{\bb{r}_m \in \RRR} \bb{f} (\bb{r}_m).
\end{equation}
Meanwhile, $\EE_{p} [\bb{f}]$ can be approximated by the expected feature values of the most likely trajectories as follows %
\begin{equation}
\label{eq:approx-exp}
\EE_{p} [\bb{f}]
\approx \bb{f} \big(\underset{\bb{r}}{\argmax}\; \log\, p (\bb{r} \,|\, \bbsym{\theta}) \big) .
\end{equation}
More specifically, for each sample trajectory $\bb{r}_m$, we fix $\bbsym{\theta}$, the trajectory $\{ \bb{x}_{1,k}, \bb{x}_{1,k+1}, \bb{u}_{1,k} \}$ of \CAV{1}, and the initial condition $\bb{x}_{2,k}$, then find the optimized control actions of \HDV{2} $\bb{u}_{2,k}$ that minimize $\bbsym{\theta}^{\top} \bb{f} (\bb{r}_m)$.
We denote the system trajectories resulted from the optimized \HDV{2}'s actions as $\{ \bb{r}_1^{\bbsym{\theta}}, \dots, \bb{r}_L^{\bbsym{\theta}} \}$.
Next, we evaluate the features for all optimized trajectories and compute the approximated expected feature values $\tilde{\EE}_{p} [\bb{f}]$  by
\begin{equation}
\label{eq:exp-fea}
\tilde{\EE}_{p} [\bb{f}]
= \frac{1}{L} \sum_{\bb{r}_m \in \RRR} \bb{f} (\bb{r}_m^{\bbsym{\theta}}).
\end{equation}

Using \eqref{eq:grad-f}, \eqref{eq:aver-fea}, and \eqref{eq:exp-fea}, the gradient of the objective function in \eqref{eq:opt-phi} with respect to $\bbsym{\theta}$ can be computed.
Therefore, the estimate of $\bbsym{\theta}$ can be updated by %
projected gradient ascent method as follows
\begin{equation}
\label{eq:pgd}
 \bbsym{\theta}^{(j+1)} =
 \mathrm{Proj}_{\Omega} \big( \bbsym{\theta}^{(j)} + \eta \nabla \LLL_{\bbsym{\theta}^{(j)}} \big) ,
\end{equation}
where $\eta \in \RRplus$ is the learning rate and $\bbsym{\theta}^{(j)}$ denotes the estimate of $\bbsym{\theta}$ at iteration $j \in \NN$ of the algorithm.

Therefore, given $L$ sample trajectories over the estimation horizon, the moving horizon IRL procedure for learning \HDV{2}'s objective weights is summarized as follows.
At each time step, we start with initial weights $\bbsym{\theta}^{(0)}$, and at each algorithmic iteration $j$, the gradient $\nabla \LLL_{\bbsym{\theta}^{(j)}}$ of the objective function in \eqref{eq:opt-phi} with respect to $\bbsym{\theta}$ at $ \bbsym{\theta} = \bbsym{\theta}^{(j)}$ is computed and used to update the estimate of $\bbsym{\theta}$ by \eqref{eq:pgd}.
For more details, the readers are referred to \cite{ziebart2008maximum} on maximum entropy IRL and to \cite{Le2022cdc} on moving horizon implementation.

\section{Optimal Weight Adaptation with Bayesian Optimization}
\label{sec:algo}
In this section, we first introduce the optimal weight adaptation problem for MPC motion planning in mixed traffic, then propose using Bayesian optimization to solve the problem.

\subsection{Optimal Weight Adaptation Problem}

Let $\bb{x}_{\text{MPC}}$ and $\bb{u}_{\text{MPC}}$ be the state and control trajectories of the agents in the simulation using MPC to control \CAV{1}.
We define the true cost in the simulation corresponding to using an MPC with a tuple of weights $\bbsym{\omega} = (\bbsym{\omega}_1, \bbsym{\omega}_2, \bbsym{\omega}_{12})$ as $J_\text{true} ^{\bbsym{\omega}} (\bb{x}_{\text{MPC}}, \bb{u}_{\text{MPC}})$.
Note that generally the true cost function $J_\text{true}$ %
can only be obtained after performing the simulations or experiments and evaluating the state and control trajectories of the agents.
We aim to seek the optimal weights of \CAV{1}'s individual objective $\bbsym{\omega}_1^{*} \in \WWW_1$ corresponding to each $(\bbsym{\omega}_2, \bbsym{\omega}_{12})$ that minimize the expected true cost given prior distribution of initial conditions $\bb{x}_{\text{MPC}} (0)$.
This can be achieved by solving the following optimization problem
\begin{equation}
\label{eq:opt-weight}
\bbsym{\omega}_1^{*} = \underset{\bbsym{\omega}_1 \in \WWW_1} \argmin \;
\bar{J}_\text{true}^{\bbsym{\omega}} (\bb{x}_{\text{MPC}}, \bb{u}_{\text{MPC}})
\end{equation}
where
\begin{equation}
\label{eq:expected-cost}
\bar{J}_\text{true}^{\bbsym{\omega}} (\bb{x}_{\text{MPC}}, \bb{u}_{\text{MPC}})
= \underset{\bb{x}_{\text{MPC}} (0)} \EE \; \big[ J_\text{true}^{\bbsym{\omega}} (\bb{x}_{\text{MPC}}, \bb{u}_{\text{MPC}}) \big] ,
\end{equation}
in which the expected true cost can be computed approximately by the average true cost of $n_s \in \NN$ independent and identically distributed
(i.i.d.) simulations with the initial states sampled from a prior distribution. %

Solving the problem in \eqref{eq:opt-weight} can be computationally intractable since the objective is a black-box function of the optimization variable $\bbsym{\omega}_1$.
Moreover, it takes a significant amount of time to evaluate that objective function because it requires multiple simulations with different initial conditions to obtain the expected true cost.
Those reasons motivate us to utilize Bayesian optimization to solve \eqref{eq:opt-weight}. %

\subsection{Bayesian Optimization}
\label{sub:bayopt}

Bayesian optimization is a machine learning-based optimization technique commonly used for minimizing (or maximizing) a black-box objective function in which we can observe only the output of the function by sampling and no first- or second-order derivatives \cite{frazier2018tutorial}.
In Bayesian optimization, the objective function is learned by a surrogate model, \eg Gaussian Process (GP), which can provide a posterior distribution of the function.
The surrogate model is combined with an acquisition function to decide the next candidate of the optimal solution.
As a result, at each algorithmic iteration, by optimizing the acquisition function over the current surrogate model, the next sampling candidate is found.
The objective value at that sampling candidate is then evaluated and added to the training data set to re-train the surrogate model.

In our problem, let $f (\bbsym{\omega}_1) = \bar{J}_\text{true}^{\bbsym{\omega}} (\bb{x}_{\text{MPC}}, \bb{u}_{\text{MPC}})$ be the black-box objective function of the variable $\bbsym{\omega}_1$ which needs to be minimized with Bayesian optimization, \ie
\begin{equation}
\underset{\bbsym{\omega}_1 \in \WWW_1} \minimize \quad f(\bbsym{\omega}_1) ,
\end{equation}
We use Gaussian process (GP) model \cite{williams2006gaussian} to learn the black-box objective function.
The GP of $f(\bbsym{\omega}_1)$ is denoted by $\GPModel_f (\bbsym{\omega}_1)$.
The GP surrogate model is combined with an acquisition function $\xi$ leading to the following optimization problem for finding the next candidate of the optimal solution
\begin{equation}
\label{eq:acquisition}
\underset{\bbsym{\omega}_1 \in \WWW_1} \maximize \quad \xi \big( \GPmean (\bbsym{\omega}_1), \GPstdvar (\bbsym{\omega}_1) \big) .
\end{equation}
where $\GPmean (\bbsym{\omega}_1)$ and $\GPstdvar (\bbsym{\omega}_1)$ denote the mean and variance of the GP prediction, respectively.
In this paper, we use the expected improvement acquisition function defined as follows
\begin{equation}
\mathrm{EI} (\bbsym{\omega}_1) = \EE \Big[ \max \{ \Delta (\bbsym{\omega}_1), 0 \} \Big],
\end{equation}
where $\Delta (\bbsym{\omega}_1) = f(\bbsym{\omega}_1^{+}) - \GPmean (\bbsym{\omega}_1)$ is the difference between the previous best sample $f(\bbsym{\omega}_1^{+})$ at $\bbsym{\omega}_1^{+}$ and the predicted output at $\bbsym{\omega}_1$.
The expected improvement under the GP model can be derived analytically as follows \cite{jones1998efficient}
\begin{equation}
\mathrm{EI}(\bbsym{\omega}_1)
= \GPstdvar (\bbsym{\omega}_1) \varphi\left(\frac{\Delta (\bbsym{\omega}_1)}{\GPstdvar (\bbsym{\omega}_1)}\right)
+ \Delta (\bbsym{\omega}_1) \Phi \left(\frac{\Delta (\bbsym{\omega}_1)}{\GPstdvar (\bbsym{\omega}_1)} \right) ,
\label{eq:EI-formula}
\end{equation}
where $\varphi$ and $\Phi$ are the probability density function (PDF) and the cumulative distribution function (CDF) of the standard normal distribution, respectively.

The entire algorithm to determine the optimal value of $\bbsym{\omega}_1$ for each $(\bbsym{\omega}_2, \bbsym{\omega}_{12})$ is summarized in Algorithm~\ref{alg:bayesopt}.
Note that we denote the candidate of the optimal solution obtained by optimizing the acquisition function at algorithmic iteration $j$ as $\bbsym{\omega}_1^{(j*)}$, which is different to the global solution $\bbsym{\omega}_1^{*}$ returned by Bayesian optimization that is the best candidate evaluated.

\begin{algorithm}[t]
  \caption{Bayesian optimization for optimal weight adaptation}
  \label{alg:bayesopt}
  \begin{algorithmic}[1]
    \Require $j_{\mathrm{max}}, j_{\mathrm{init}} \in \NN$, \(\bbsym{\omega}_2, \bbsym{\omega}_{12}\)
    \Procedure{Initialization}{}
    \For{$j = 1,2,\dots,j_{\mathrm{init}}$}
    \State Randomly sample $\bbsym{\omega}_1^{(j)} \in \WWW_1$
    \State Compute average true cost $\bar{J}_\text{true}^{\bbsym{\omega}^{(j)}}$ %
    \State Add $(\bbsym{\omega}_1^{(j)}, \bar{J}_\text{true}^{\bbsym{\omega}^{(j)}})$ to a training dataset $\DDD$
    \EndFor
    \State Learn a GP model $\GPModel (\bbsym{\omega}_1)$ with $\DDD$
    \EndProcedure
    \Procedure{Bayesian Optimization}{}
    \For {$j = 1, \dots, j_{\mathrm{max}}$}
    \State Find next candidate \(\bbsym{\omega}_1^{(j*)}\) by optimizing the acquisition function %
    \State Compute average true cost $\bar{J}_\text{true}^{\bbsym{\omega}^{(j*)}}$ %
    \State Add $(\bbsym{\omega}_1^{(j*)}, \bar{J}_\text{true}^{\bbsym{\omega}^{(j*)}})$ to $\DDD$ and re-train $\GPModel (\bbsym{\omega}_1)$
    \EndFor 
    \State \textbf{return} $\bbsym{\omega}_1^{*}$
    \EndProcedure
  \end{algorithmic}
\end{algorithm} \setlength{\textfloatsep}{0.05cm}

\section{Illustrative Example}

\begin{figure}[!t]
\centering
    \includegraphics[scale=0.32, bb = 150 20 850 520, clip=true]{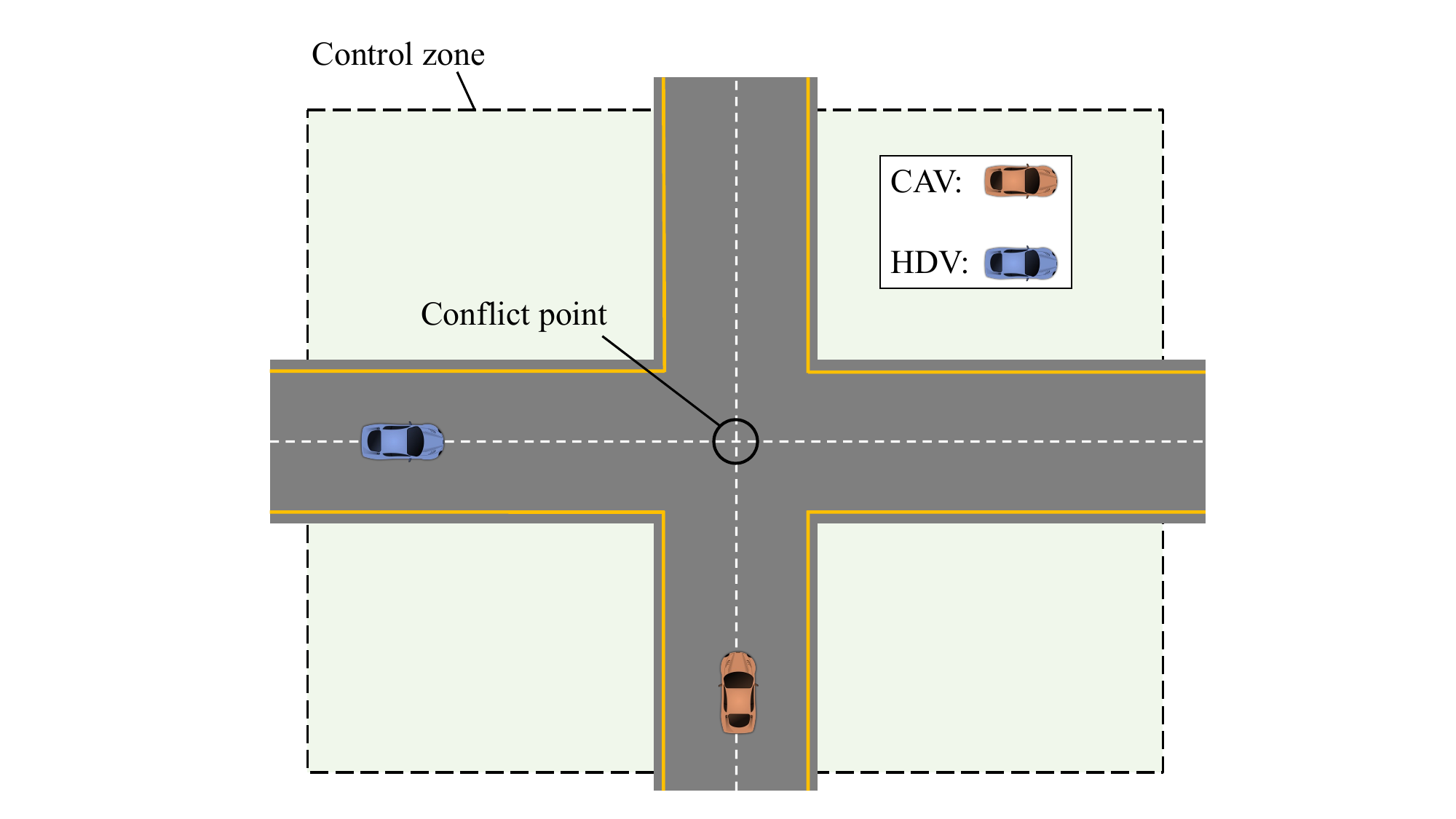}
    \caption{An unsignalized intersection scenario with a CAV and an HDV.}
    \label{fig:intersection}
    \vspace{-5pt}
\end{figure}

In this section, we demonstrate the control formulation presented in Section~\ref{sec:mpc} and the optimal weight adaptation problem in Section~\ref{sec:algo} by a vehicle crossing example at an unsignalized intersection illustrated in Fig.~\ref{fig:intersection}.
We define the surrounding area of the intersection inside of which the vehicles can communicate with the coordinator as a control zone, while the location where a lateral collision can occur is called a conflict point. 
The dynamics of each vehicle $i$ are described by the following double-integrator dynamics
\begin{equation}
\label{eq:integrator}
\begin{split}
p_{i,k+1} &= p_{i,k} + \Delta T v_{i,k} + \frac{1}{2} \Delta T^2 a_{i,k} , \\
v_{i,k+1} &= v_{i,k} + \Delta T a_{i,k} , \\
\end{split}
\end{equation}
where $\Delta T \in \RRplus$ is the sampling time, $p_{i,k} \in \RR$ is the longitudinal position of the vehicle with respect to the conflict point at time $k$, and  $v_{i,k} \in \RR$ and $a_{i,k} \in \RR$ are the speed and acceleration of the vehicle $i$ at time $k$, respectively.
The state and control input of vehicle $i$ are defined by $\bb{x}_{i,k} = [p_{i,k}, v_{i,k}]^{\top}$ and $u_{i,k} = a_{i,k}$, respectively.

The individual objective for each vehicle in the MPC problem includes: (1) minimizing the control input for smoother movement and energy saving,
and (2) minimizing the deviation from the maximum allowed speed to reduce the time to cross the intersection, \ie
\begin{equation}
\label{eq:ex-ego1}
l_{i} (\bb{x}_{i,k+1}, u_{i,k})
= \begin{bmatrix}
\omega_{i,1} \\
\omega_{i,2}
\end{bmatrix}^\top
\begin{bmatrix}
a_{i,k}^2 \\
(v_{i,k+1} - v_{\text{max}})^2
\end{bmatrix} ,
\end{equation}
for $i = 1,2$, where $\omega_{i,1}$, $\omega_{i,2} \in \RRplus$ are positive weights.
The shared objective function takes the form of a logarithmic penalty function corresponding to a collision avoidance constraint %
as follows
\begin{equation}
\label{eq:ex-cooperative}
l_{12} (\bb{x}_{1,k+1}, \bb{x}_{2,k+1})
= - \omega_{12} \log \big( \gamma (p_{1,k+1}^2 + p_{2,k+1}^2) \big) ,
\end{equation}
where $\omega_{12} \in \RRplus$ is a positive weight and and $\gamma \in \RRplus$ is a parameter of the logarithmic penalty function.

Next, we consider the following state and control constraints for \CAV{1}
\begin{equation}
\label{eq:bound}
v_{\text{min}} \le v_{1,k+1} \le v_{\text{max}},\; u_{\text{min}} \le a_{1,k} \le u_{\text{max}}, \; \forall k \in \III_t,
\end{equation}
where $u_{\text{min}}$, $u_{\text{max}} \in \RR$ are the minimum deceleration and maximum acceleration, respectively, and $v_{\text{min}}$, $v_{\text{max}} \in \RR$ are the minimum and maximum speed limits, respectively.
Moreover, we impose the following safety constraint
\begin{equation}
\label{eq:safe-dist}
r \le \sqrt{p_{1,k+1}^2 + p_{2,k+1}^2}, \; \forall k \in \III_t,
\end{equation}
to guarantee that the predicted distances between \CAV{1} and \HDV{2} are greater than a safety threshold $r \in \RRplus$.

The MPC problem for \CAV{1} in this example is thus formulated as follows
\begin{subequations}
  \label{eq:ex-mpc}
  \begin{align}
    &
    \begin{multlined}
    \underset{ \{u_{1,k}, u_{2,k}\}_{k \in \III_t} }{\minimize} \; \sum_{k \in \III_t} \Big( \sum_{i = 1,2} l_{i} (\bb{x}_{i,k+1}, u_{i,k}) \\ 
    + l_{12} (\bb{x}_{1,k+1}, \bb{x}_{2,k+1}) \Big),
    \end{multlined}
    \label{eq:ex-mpc:obj}\\
    & \text{subject to:} \nonumber  \\
    & \quad \text{\eqref{eq:integrator}},\; \forall k \in \III_t, \, i = 1,2, \label{eq:ex-mpc-dyn} \\
    & \quad \text{\eqref{eq:bound}}, \text{\eqref{eq:safe-dist}}, \; \forall k \in \III_t, \label{eq:ex-mpc-bound} %
  \end{align}
\end{subequations}

\label{sec:example}

We define a true cost function called \emph{time-energy efficiency with safety} that is computed by %
\begin{equation}
\label{eq:ex-true-cost}
J_\text{true}^{\bbsym{\omega}} (\bb{x}_{\text{MPC}}, \bb{u}_{\text{MPC}}) = \alpha t_{1,f} + \beta E_t + \lambda \mathbb{I} \big( g(\bb{x}_{\text{MPC}}) \big),
\end{equation}
where $\alpha$, $\beta$, and $\lambda \in \RRplus$ are constant weights and $\lambda$ is sufficiently large compared to $\alpha$ and $\beta$ to prioritize safety rather than time and energy efficiency,
$t_{1,f}$ is the time that \CAV{1} exits the control zone, $E$ is the total amount of energy consumption of \CAV{1} %
while traveling in the control zone, and $\mathbb{I} \big( g(\bb{x}_{\text{MPC}}))$ is the indicator function of the safety constraint $g$ defined as
\begin{equation}
\label{eq:safety-indicator}
\mathbb{I} \big( g(\bb{x}_{\text{MPC}}) \big) =
\begin{cases}
  0, & \text{if}\ g(\bb{x}_{\text{MPC}}) \le 0  \\
  1, & \text{otherwise}
\end{cases}.
\end{equation}
The safety constraint is $g(\bb{x}_{\text{MPC}}) = r - d_{12, \text{min}} \le 0$
where $d_{12, \text{min}}$ is the minimum distance between two vehicles.
Within Bayesian optimization framework that requires a continuous objective function, we approximate the indicator function by a sigmoid function. %
To evaluate the total fuel consumption $E_t$ of \CAV{1}, we consider the polynomial meta-model and coefficients from an engine torque-speed-efficiency map of a typical car presented in \cite{kamal2012model}.

\section{Simulation Results}
\label{sec:sim}
To demonstrate the effectiveness of the proposed method, we conduct numerical simulations for the intersection crossing example described in Section~\ref{sec:example}.

\subsection{Simulation Setup}

\begin{figure}[!tb]
\centering
\begin{subfigure}{.48\textwidth}
\centering
\scalebox{.85}{\input{figs/colormap_w1.tex}}
\caption{Heat map for $\log_{10} (\omega_{1,1})$}
\end{subfigure}
\begin{subfigure}{.48\textwidth}
\centering
\scalebox{.85}{\input{figs/colormap_w2.tex}}
\caption{Heat map for $\log_{10} (\omega_{1,2})$}
\end{subfigure}
\caption{Heat maps for the optimal weight adaptation strategy (in log scale). The black crosses represent the sampled values of $\log_{10} (\bbsym{\omega}_{2})$.}
\label{fig:colormap}
\vspace{-5pt}
\end{figure}
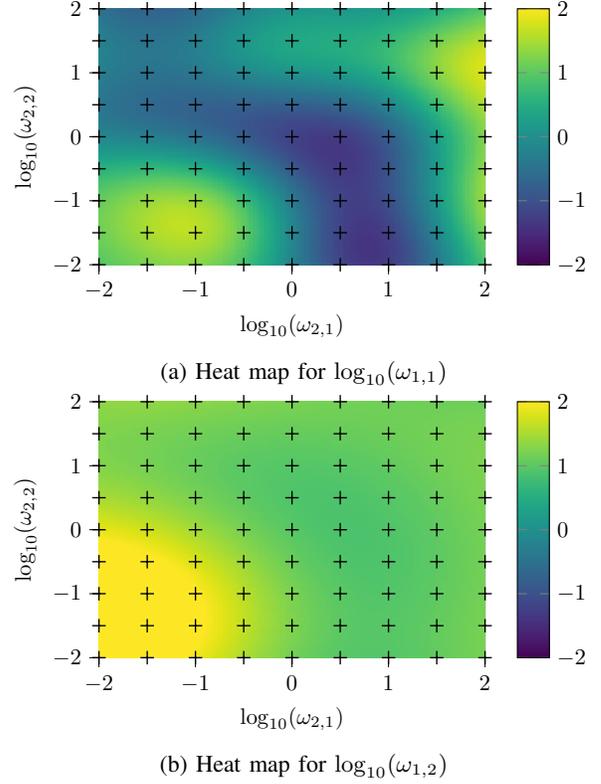

For the implementation, since the solution of the MPC problem does not change if all the weights are scaled by a positive factor, 
we fix the shared objective weight $\omega_{12} = 10^3$ to reduce the dimension of the problem. 
We consider $\WWW_i = \{ \omega_{i,1}, \omega_{i,2} \; | \; 10^{-2} \le \omega_{i,1}, \omega_{i,2} \le 10^2 \} $, for $i = 1,2$ and create a grid of size $9\times 9$ linearly spaced in log scale for $\bbsym{\omega}_{2}$.
For each $\bbsym{\omega}_{2}$ in the grid, we employ Bayesian optimization to find the optimal value for $\bbsym{\omega}_{1}$.
The average true cost of MPC is computed by averaging the true cost values in $n_s = 100$ i.i.d. simulations with uniformly distributed initial positions and velocities.   
The parameters in the Bayesian optimization algorithm and in the true cost are chosen as $j_\mathrm{max} = 25$, $j_\mathrm{init} = 5$, $\alpha = 1.0$, $\beta = 1.0$, $\lambda = 10^3$, $r = \SI{10.0}{(m)}$, $\xi = 10.0$. 
The grid and corresponding solutions are then used as training data for GP regression to learn the weight adaptation strategy. %
The derived optimal weight adaptation strategy can be illustrated by heat maps in Fig.~\ref{fig:colormap}.

In the testing simulations, we generate the actions of the human drivers by using the solution of \eqref{eq:utility} in which the weights are varied to imitate different driving behavior.
Note that in all the simulations the \HDV{2}'s objective weights are unknown %
to \CAV{1} and must be learned online by moving horizon IRL.
The parameters of MPC and moving horizon IRL are chosen as: 
$\Delta T = \SI{0.2}{s}$, $H = 10$, $\gamma = 1.0$, $v_\mathrm{min} = \SI{0.0}{(m/s)}$, $v_\mathrm{max} = \SI{12.0}{(m/s)}$, $u_\mathrm{min} = \SI{-5.0}{(m/s^2)}$, $u_\mathrm{max} = \SI{3.0}{(m/s^2)}$, $L = 20$, $\eta = 0.01$.
The simulation is implemented in Julia programming language, and %
KNITRO solver \cite{byrd2006k} is used for solving MPC problems. %
The code for simulations is available online at \url{https://github.com/vietanhle0101/MPC-BayesOpt-Mixed-Traffic}.

\subsection{Results and Discussion}

Using the obtained strategy for MPC weight adaptation, we first evaluate the control framework in two specific simulations with an altruistic driver and with an egoistic driver to demonstrate that \CAV{1} behaves differently depending on human driving behavior. 
The video for those simulations can be found in \url{https://sites.google.com/view/ud-ids-lab/mpc-bayesopt}.

\begin{figure}[!tb]
\centering
\scalebox{0.9}{\input{figs/histogram.tex}}
\vspace{-5pt}
\caption{A histogram for percentages of improvement in 5000 simulations.}
\label{fig:boxplot}
\vspace{-5pt}
\end{figure}
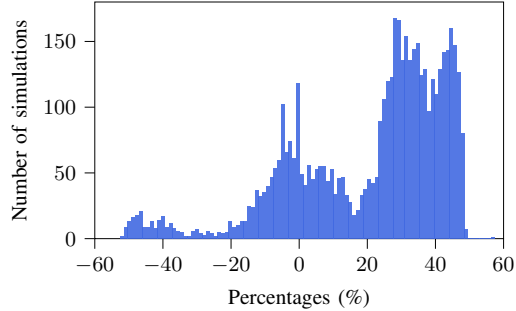

\textbf{Comparison with a baseline strategy:} 
We compare the performance of the optimal weight adaptation strategy with a baseline strategy using SVO \cite{Le2022cdc}. 
To extensively assess the benefits of the proposed method, we conduct $5000$ simulations with different initial conditions of the vehicles and heterogeneous driving styles of the human drivers. 
First, we compare %
by two metrics: (1) the number of simulations without unsafe situations, and (2) the number of simulations with time-energy improvement among all the configurations in which using both strategies do not cause unsafe situations, as indicated in Table~\ref{tab:compare}.
It can be observed that with a roughly similar level of safety (higher than $99 \%$), MPC weight adaptation with the optimal strategy performs better than with the socially cooperative strategy in approximately 80\% of the simulations.
Furthermore, we also compute the percentages of improvement in time-energy costs 
and show the results in a histogram form in Fig.~\ref{fig:boxplot}.
We have been able to improve the average performance by $20.7 \%$. %

\begin{table}[!bt]
  \caption{Comparison between weight adaptation strategies using Bayesian optimization (BayOpt) and SVO.}
  \label{tab:compare} 
  \centering
  \begin{tabular}{ p{0.22\textwidth} | p{0.095\textwidth} p{0.095\textwidth} }
    \toprule[1pt]%
    \textbf{Comparison metrics} & \textbf{BayOpt} & \textbf{SVO} \\
    \midrule[0.5pt] %
    Number of simulations with safety & $4983$ ($99.7 \%$) & $4981$ ($99.6 \%$) \\
    Number of simulations with time-energy improvement\footnotemark & $3921$ ($79.0 \%$) & $1043$ ($21.0 \%$)\\
    \bottomrule[1pt] %
  \end{tabular}
  \vspace{-5pt}
\end{table}
\footnotetext{Among all the configurations in which using both strategies can avoid unsafe situations.}

\section{Conclusions}
\label{sec:conc}
In this paper, we presented a method to derive an optimal weight adaptation strategy of MPC for CAVs in mixed traffic with Bayesian optimization.
By numerical simulations of a vehicle crossing example at an unsignalized intersection, we showed that the proposed optimal weight adaptation strategy has approximately $20\%$ improvement on average over a baseline strategy using social value orientation.
As a future research direction, we plan to focus on (1) enhancing the framework with a safety-guarantee mechanism, and (2) validating it in an experimental testbed \cite{chalaki2021CSM}.

\bibliographystyle{IEEEtran}
\bibliography{IEEEabrv,references,references_IDS}

\balance

\end{document}

%% file: defs.tex
\newcommand{\GPModel}{\ensuremath{\GGG}}

\newcommand{\GPmean}{\ensuremath{\mu}}

\newcommand{\GPstdvar}{\ensuremath{\sigma}}

\newcommand{\bb}[1]{\ensuremath{\mathbf{#1}}}
\newcommand{\bbsym}[1]{\ensuremath{\boldsymbol{#1}}}
\newcommand{\CAV}[1]{CAV\textendash \ensuremath{#1}\xspace}
\newcommand{\HDV}[1]{HDV\textendash\ensuremath{#1}\xspace}
\newcommand{\hrm}[1]{\ensuremath{\hspace{-#1pt}}}

%% file: figs/colormap_w1.tex
\begin{tikzpicture}

\definecolor{darkgray176}{RGB}{176,176,176}

\begin{axis}[
width = 3in, height = 2.2in,
colorbar,
colorbar style={ylabel={}},
colormap/viridis,
point meta max=2,
point meta min=-2,
tick align=outside,
tick pos=left,
x grid style={darkgray176},
xmin=-2, xmax=2,
xlabel={$\log_{10} (\omega_{2,1})$},
xtick style={color=black},
y grid style={darkgray176},
ymin=-2, ymax=2,
ylabel={$\log_{10} (\omega_{2,2})$},
ytick style={color=black}
]
\addplot graphics [includegraphics cmd=\pgfimage,xmin=-2, xmax=2, ymin=-2, ymax=2] {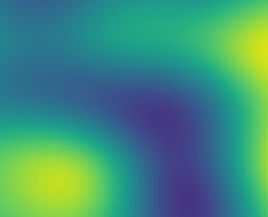};
\addplot [semithick, black, mark=+, mark size=3, mark options={solid}, only marks]
table {%
-2 -2
-1.5 -2
-1 -2
-0.5 -2
0 -2
0.5 -2
1 -2
1.5 -2
2 -2
-2 -1.5
-1.5 -1.5
-1 -1.5
-0.5 -1.5
0 -1.5
0.5 -1.5
1 -1.5
1.5 -1.5
2 -1.5
-2 -1
-1.5 -1
-1 -1
-0.5 -1
0 -1
0.5 -1
1 -1
1.5 -1
2 -1
-2 -0.5
-1.5 -0.5
-1 -0.5
-0.5 -0.5
0 -0.5
0.5 -0.5
1 -0.5
1.5 -0.5
2 -0.5
-2 0
-1.5 0
-1 0
-0.5 0
0 0
0.5 0
1 0
1.5 0
2 0
-2 0.5
-1.5 0.5
-1 0.5
-0.5 0.5
0 0.5
0.5 0.5
1 0.5
1.5 0.5
2 0.5
-2 1
-1.5 1
-1 1
-0.5 1
0 1
0.5 1
1 1
1.5 1
2 1
-2 1.5
-1.5 1.5
-1 1.5
-0.5 1.5
0 1.5
0.5 1.5
1 1.5
1.5 1.5
2 1.5
-2 2
-1.5 2
-1 2
-0.5 2
0 2
0.5 2
1 2
1.5 2
2 2
};
\end{axis}

\end{tikzpicture}

%% file: figs/colormap_w2.tex
\begin{tikzpicture}

\definecolor{darkgray176}{RGB}{176,176,176}

\begin{axis}[
width = 3.in, height = 2.2in,
colorbar,
colorbar style={ylabel={}},
colormap/viridis,
point meta max=2,
point meta min=-2,
tick align=outside,
tick pos=left,
x grid style={darkgray176},
xmin=-2, xmax=2,
xlabel={$\log_{10} (\omega_{2,1})$},
xtick style={color=black},
y grid style={darkgray176},
ymin=-2, ymax=2,
ylabel={$\log_{10} (\omega_{2,2})$},
ytick style={color=black}
]
\addplot graphics [includegraphics cmd=\pgfimage,xmin=-2, xmax=2, ymin=-2, ymax=2] {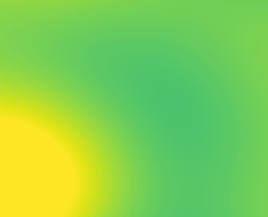};
\addplot [semithick, black, mark=+, mark size=3, mark options={solid}, only marks]
table {%
-2 -2
-1.5 -2
-1 -2
-0.5 -2
0 -2
0.5 -2
1 -2
1.5 -2
2 -2
-2 -1.5
-1.5 -1.5
-1 -1.5
-0.5 -1.5
0 -1.5
0.5 -1.5
1 -1.5
1.5 -1.5
2 -1.5
-2 -1
-1.5 -1
-1 -1
-0.5 -1
0 -1
0.5 -1
1 -1
1.5 -1
2 -1
-2 -0.5
-1.5 -0.5
-1 -0.5
-0.5 -0.5
0 -0.5
0.5 -0.5
1 -0.5
1.5 -0.5
2 -0.5
-2 0
-1.5 0
-1 0
-0.5 0
0 0
0.5 0
1 0
1.5 0
2 0
-2 0.5
-1.5 0.5
-1 0.5
-0.5 0.5
0 0.5
0.5 0.5
1 0.5
1.5 0.5
2 0.5
-2 1
-1.5 1
-1 1
-0.5 1
0 1
0.5 1
1 1
1.5 1
2 1
-2 1.5
-1.5 1.5
-1 1.5
-0.5 1.5
0 1.5
0.5 1.5
1 1.5
1.5 1.5
2 1.5
-2 2
-1.5 2
-1 2
-0.5 2
0 2
0.5 2
1 2
1.5 2
2 2
};
\end{axis}

\end{tikzpicture}

%% file: figs/histogram.tex
\begin{tikzpicture}[font=\small]

\definecolor{darkgray176}{RGB}{65,105,225}
\definecolor{lightblue}{RGB}{173,216,230}
\definecolor{deepskyblue}{RGB}{65,105,225}

\begin{axis}[
width = 3.in, height = 2in,
tick align=outside,
tick pos=left,
x grid style={darkgray176},
xlabel={Percentages (\%)},
xmin=-60.0, xmax=60.0,
xtick style={color=black},
y grid style={darkgray176},
ylabel={Number of simulations},
ymin=0, ymax=180.0,
ytick style={color=black},
ytick = {0,50,100,150},
]
\draw[draw=none,fill=deepskyblue,fill opacity=0.9] (axis cs:-52.5916072909825,0) rectangle (axis cs:-51.4924191164248,2);
\draw[draw=none,fill=deepskyblue,fill opacity=0.9] (axis cs:-51.4924191164248,0) rectangle (axis cs:-50.3932309418672,9);
\draw[draw=none,fill=deepskyblue,fill opacity=0.9] (axis cs:-50.3932309418672,0) rectangle (axis cs:-49.2940427673096,13);
\draw[draw=none,fill=deepskyblue,fill opacity=0.9] (axis cs:-49.2940427673096,0) rectangle (axis cs:-48.1948545927519,16);
\draw[draw=none,fill=deepskyblue,fill opacity=0.9] (axis cs:-48.1948545927519,0) rectangle (axis cs:-47.0956664181943,18);
\draw[draw=none,fill=deepskyblue,fill opacity=0.9] (axis cs:-47.0956664181943,0) rectangle (axis cs:-45.9964782436367,21);
\draw[draw=none,fill=deepskyblue,fill opacity=0.9] (axis cs:-45.9964782436367,0) rectangle (axis cs:-44.8972900690791,9);
\draw[draw=none,fill=deepskyblue,fill opacity=0.9] (axis cs:-44.8972900690791,0) rectangle (axis cs:-43.7981018945214,9);
\draw[draw=none,fill=deepskyblue,fill opacity=0.9] (axis cs:-43.7981018945214,0) rectangle (axis cs:-42.6989137199638,13);
\draw[draw=none,fill=deepskyblue,fill opacity=0.9] (axis cs:-42.6989137199638,0) rectangle (axis cs:-41.5997255454062,8);
\draw[draw=none,fill=deepskyblue,fill opacity=0.9] (axis cs:-41.5997255454062,0) rectangle (axis cs:-40.5005373708485,14);
\draw[draw=none,fill=deepskyblue,fill opacity=0.9] (axis cs:-40.5005373708485,0) rectangle (axis cs:-39.4013491962909,17);
\draw[draw=none,fill=deepskyblue,fill opacity=0.9] (axis cs:-39.4013491962909,0) rectangle (axis cs:-38.3021610217333,9);
\draw[draw=none,fill=deepskyblue,fill opacity=0.9] (axis cs:-38.3021610217333,0) rectangle (axis cs:-37.2029728471757,12);
\draw[draw=none,fill=deepskyblue,fill opacity=0.9] (axis cs:-37.2029728471757,0) rectangle (axis cs:-36.103784672618,8);
\draw[draw=none,fill=deepskyblue,fill opacity=0.9] (axis cs:-36.103784672618,0) rectangle (axis cs:-35.0045964980604,6);
\draw[draw=none,fill=deepskyblue,fill opacity=0.9] (axis cs:-35.0045964980604,0) rectangle (axis cs:-33.9054083235028,5);
\draw[draw=none,fill=deepskyblue,fill opacity=0.9] (axis cs:-33.9054083235028,0) rectangle (axis cs:-32.8062201489451,2);
\draw[draw=none,fill=deepskyblue,fill opacity=0.9] (axis cs:-32.8062201489451,0) rectangle (axis cs:-31.7070319743875,2);
\draw[draw=none,fill=deepskyblue,fill opacity=0.9] (axis cs:-31.7070319743875,0) rectangle (axis cs:-30.6078437998299,6);
\draw[draw=none,fill=deepskyblue,fill opacity=0.9] (axis cs:-30.6078437998299,0) rectangle (axis cs:-29.5086556252722,7);
\draw[draw=none,fill=deepskyblue,fill opacity=0.9] (axis cs:-29.5086556252722,0) rectangle (axis cs:-28.4094674507146,4);
\draw[draw=none,fill=deepskyblue,fill opacity=0.9] (axis cs:-28.4094674507146,0) rectangle (axis cs:-27.310279276157,3);
\draw[draw=none,fill=deepskyblue,fill opacity=0.9] (axis cs:-27.310279276157,0) rectangle (axis cs:-26.2110911015994,6);
\draw[draw=none,fill=deepskyblue,fill opacity=0.9] (axis cs:-26.2110911015994,0) rectangle (axis cs:-25.1119029270417,4);
\draw[draw=none,fill=deepskyblue,fill opacity=0.9] (axis cs:-25.1119029270417,0) rectangle (axis cs:-24.0127147524841,2);
\draw[draw=none,fill=deepskyblue,fill opacity=0.9] (axis cs:-24.0127147524841,0) rectangle (axis cs:-22.9135265779265,5);
\draw[draw=none,fill=deepskyblue,fill opacity=0.9] (axis cs:-22.9135265779265,0) rectangle (axis cs:-21.8143384033688,4);
\draw[draw=none,fill=deepskyblue,fill opacity=0.9] (axis cs:-21.8143384033688,0) rectangle (axis cs:-20.7151502288112,5);
\draw[draw=none,fill=deepskyblue,fill opacity=0.9] (axis cs:-20.7151502288112,0) rectangle (axis cs:-19.6159620542536,13);
\draw[draw=none,fill=deepskyblue,fill opacity=0.9] (axis cs:-19.6159620542536,0) rectangle (axis cs:-18.5167738796959,9);
\draw[draw=none,fill=deepskyblue,fill opacity=0.9] (axis cs:-18.5167738796959,0) rectangle (axis cs:-17.4175857051383,10);
\draw[draw=none,fill=deepskyblue,fill opacity=0.9] (axis cs:-17.4175857051383,0) rectangle (axis cs:-16.3183975305807,13);
\draw[draw=none,fill=deepskyblue,fill opacity=0.9] (axis cs:-16.3183975305807,0) rectangle (axis cs:-15.2192093560231,13);
\draw[draw=none,fill=deepskyblue,fill opacity=0.9] (axis cs:-15.2192093560231,0) rectangle (axis cs:-14.1200211814654,25);
\draw[draw=none,fill=deepskyblue,fill opacity=0.9] (axis cs:-14.1200211814654,0) rectangle (axis cs:-13.0208330069078,24);
\draw[draw=none,fill=deepskyblue,fill opacity=0.9] (axis cs:-13.0208330069078,0) rectangle (axis cs:-11.9216448323502,37);
\draw[draw=none,fill=deepskyblue,fill opacity=0.9] (axis cs:-11.9216448323502,0) rectangle (axis cs:-10.8224566577925,32);
\draw[draw=none,fill=deepskyblue,fill opacity=0.9] (axis cs:-10.8224566577925,0) rectangle (axis cs:-9.7232684832349,35);
\draw[draw=none,fill=deepskyblue,fill opacity=0.9] (axis cs:-9.7232684832349,0) rectangle (axis cs:-8.62408030867728,40);
\draw[draw=none,fill=deepskyblue,fill opacity=0.9] (axis cs:-8.62408030867728,0) rectangle (axis cs:-7.52489213411965,47);
\draw[draw=none,fill=deepskyblue,fill opacity=0.9] (axis cs:-7.52489213411965,0) rectangle (axis cs:-6.42570395956201,54);
\draw[draw=none,fill=deepskyblue,fill opacity=0.9] (axis cs:-6.42570395956201,0) rectangle (axis cs:-5.32651578500439,60);
\draw[draw=none,fill=deepskyblue,fill opacity=0.9] (axis cs:-5.32651578500439,0) rectangle (axis cs:-4.22732761044676,102);
\draw[draw=none,fill=deepskyblue,fill opacity=0.9] (axis cs:-4.22732761044676,0) rectangle (axis cs:-3.12813943588912,66);
\draw[draw=none,fill=deepskyblue,fill opacity=0.9] (axis cs:-3.12813943588912,0) rectangle (axis cs:-2.0289512613315,74);
\draw[draw=none,fill=deepskyblue,fill opacity=0.9] (axis cs:-2.0289512613315,0) rectangle (axis cs:-0.929763086773868,61);
\draw[draw=none,fill=deepskyblue,fill opacity=0.9] (axis cs:-0.929763086773868,0) rectangle (axis cs:0.169425087783765,118);
\draw[draw=none,fill=deepskyblue,fill opacity=0.9] (axis cs:0.169425087783765,0) rectangle (axis cs:1.26861326234139,49);
\draw[draw=none,fill=deepskyblue,fill opacity=0.9] (axis cs:1.26861326234139,0) rectangle (axis cs:2.36780143689902,41);
\draw[draw=none,fill=deepskyblue,fill opacity=0.9] (axis cs:2.36780143689902,0) rectangle (axis cs:3.46698961145665,56);
\draw[draw=none,fill=deepskyblue,fill opacity=0.9] (axis cs:3.46698961145665,0) rectangle (axis cs:4.56617778601428,45);
\draw[draw=none,fill=deepskyblue,fill opacity=0.9] (axis cs:4.56617778601428,0) rectangle (axis cs:5.66536596057191,53);
\draw[draw=none,fill=deepskyblue,fill opacity=0.9] (axis cs:5.66536596057191,0) rectangle (axis cs:6.76455413512954,55);
\draw[draw=none,fill=deepskyblue,fill opacity=0.9] (axis cs:6.76455413512954,0) rectangle (axis cs:7.86374230968717,55);
\draw[draw=none,fill=deepskyblue,fill opacity=0.9] (axis cs:7.86374230968717,0) rectangle (axis cs:8.9629304842448,44);
\draw[draw=none,fill=deepskyblue,fill opacity=0.9] (axis cs:8.9629304842448,0) rectangle (axis cs:10.0621186588024,53);
\draw[draw=none,fill=deepskyblue,fill opacity=0.9] (axis cs:10.0621186588024,0) rectangle (axis cs:11.1613068333601,34);
\draw[draw=none,fill=deepskyblue,fill opacity=0.9] (axis cs:11.1613068333601,0) rectangle (axis cs:12.2604950079177,46);
\draw[draw=none,fill=deepskyblue,fill opacity=0.9] (axis cs:12.2604950079177,0) rectangle (axis cs:13.3596831824753,47);
\draw[draw=none,fill=deepskyblue,fill opacity=0.9] (axis cs:13.3596831824753,0) rectangle (axis cs:14.4588713570329,33);
\draw[draw=none,fill=deepskyblue,fill opacity=0.9] (axis cs:14.4588713570329,0) rectangle (axis cs:15.5580595315906,28);
\draw[draw=none,fill=deepskyblue,fill opacity=0.9] (axis cs:15.5580595315906,0) rectangle (axis cs:16.6572477061482,18);
\draw[draw=none,fill=deepskyblue,fill opacity=0.9] (axis cs:16.6572477061482,0) rectangle (axis cs:17.7564358807058,22);
\draw[draw=none,fill=deepskyblue,fill opacity=0.9] (axis cs:17.7564358807058,0) rectangle (axis cs:18.8556240552635,33);
\draw[draw=none,fill=deepskyblue,fill opacity=0.9] (axis cs:18.8556240552635,0) rectangle (axis cs:19.9548122298211,38);
\draw[draw=none,fill=deepskyblue,fill opacity=0.9] (axis cs:19.9548122298211,0) rectangle (axis cs:21.0540004043787,45);
\draw[draw=none,fill=deepskyblue,fill opacity=0.9] (axis cs:21.0540004043787,0) rectangle (axis cs:22.1531885789364,42);
\draw[draw=none,fill=deepskyblue,fill opacity=0.9] (axis cs:22.1531885789364,0) rectangle (axis cs:23.252376753494,47);
\draw[draw=none,fill=deepskyblue,fill opacity=0.9] (axis cs:23.252376753494,0) rectangle (axis cs:24.3515649280516,89);
\draw[draw=none,fill=deepskyblue,fill opacity=0.9] (axis cs:24.3515649280516,0) rectangle (axis cs:25.4507531026092,106);
\draw[draw=none,fill=deepskyblue,fill opacity=0.9] (axis cs:25.4507531026092,0) rectangle (axis cs:26.5499412771669,120);
\draw[draw=none,fill=deepskyblue,fill opacity=0.9] (axis cs:26.5499412771669,0) rectangle (axis cs:27.6491294517245,123);
\draw[draw=none,fill=deepskyblue,fill opacity=0.9] (axis cs:27.6491294517245,0) rectangle (axis cs:28.7483176262821,168);
\draw[draw=none,fill=deepskyblue,fill opacity=0.9] (axis cs:28.7483176262821,0) rectangle (axis cs:29.8475058008398,166);
\draw[draw=none,fill=deepskyblue,fill opacity=0.9] (axis cs:29.8475058008398,0) rectangle (axis cs:30.9466939753974,136);
\draw[draw=none,fill=deepskyblue,fill opacity=0.9] (axis cs:30.9466939753974,0) rectangle (axis cs:32.045882149955,154);
\draw[draw=none,fill=deepskyblue,fill opacity=0.9] (axis cs:32.045882149955,0) rectangle (axis cs:33.1450703245127,136);
\draw[draw=none,fill=deepskyblue,fill opacity=0.9] (axis cs:33.1450703245127,0) rectangle (axis cs:34.2442584990703,144);
\draw[draw=none,fill=deepskyblue,fill opacity=0.9] (axis cs:34.2442584990703,0) rectangle (axis cs:35.3434466736279,149);
\draw[draw=none,fill=deepskyblue,fill opacity=0.9] (axis cs:35.3434466736279,0) rectangle (axis cs:36.4426348481855,124);
\draw[draw=none,fill=deepskyblue,fill opacity=0.9] (axis cs:36.4426348481855,0) rectangle (axis cs:37.5418230227432,129);
\draw[draw=none,fill=deepskyblue,fill opacity=0.9] (axis cs:37.5418230227432,0) rectangle (axis cs:38.6410111973008,97);
\draw[draw=none,fill=deepskyblue,fill opacity=0.9] (axis cs:38.6410111973008,0) rectangle (axis cs:39.7401993718584,121);
\draw[draw=none,fill=deepskyblue,fill opacity=0.9] (axis cs:39.7401993718584,0) rectangle (axis cs:40.8393875464161,110);
\draw[draw=none,fill=deepskyblue,fill opacity=0.9] (axis cs:40.8393875464161,0) rectangle (axis cs:41.9385757209737,129);
\draw[draw=none,fill=deepskyblue,fill opacity=0.9] (axis cs:41.9385757209737,0) rectangle (axis cs:43.0377638955313,142);
\draw[draw=none,fill=deepskyblue,fill opacity=0.9] (axis cs:43.0377638955313,0) rectangle (axis cs:44.136952070089,143);
\draw[draw=none,fill=deepskyblue,fill opacity=0.9] (axis cs:44.136952070089,0) rectangle (axis cs:45.2361402446466,160);
\draw[draw=none,fill=deepskyblue,fill opacity=0.9] (axis cs:45.2361402446466,0) rectangle (axis cs:46.3353284192042,147);
\draw[draw=none,fill=deepskyblue,fill opacity=0.9] (axis cs:46.3353284192042,0) rectangle (axis cs:47.4345165937618,127);
\draw[draw=none,fill=deepskyblue,fill opacity=0.9] (axis cs:47.4345165937618,0) rectangle (axis cs:48.5337047683195,80);
\draw[draw=none,fill=deepskyblue,fill opacity=0.9] (axis cs:48.5337047683195,0) rectangle (axis cs:49.6328929428771,7);
\draw[draw=none,fill=deepskyblue,fill opacity=0.9] (axis cs:49.6328929428771,0) rectangle (axis cs:50.7320811174347,0);
\draw[draw=none,fill=deepskyblue,fill opacity=0.9] (axis cs:50.7320811174347,0) rectangle (axis cs:51.8312692919924,0);
\draw[draw=none,fill=deepskyblue,fill opacity=0.9] (axis cs:51.8312692919924,0) rectangle (axis cs:52.93045746655,0);
\draw[draw=none,fill=deepskyblue,fill opacity=0.9] (axis cs:52.93045746655,0) rectangle (axis cs:54.0296456411076,0);
\draw[draw=none,fill=deepskyblue,fill opacity=0.9] (axis cs:54.0296456411076,0) rectangle (axis cs:55.1288338156652,0);
\draw[draw=none,fill=deepskyblue,fill opacity=0.9] (axis cs:55.1288338156652,0) rectangle (axis cs:56.2280219902229,0);
\draw[draw=none,fill=deepskyblue,fill opacity=0.9] (axis cs:56.2280219902229,0) rectangle (axis cs:57.3272101647805,1);
\end{axis}

\end{tikzpicture}